\documentclass[twocolumn,aps,pra,amsmath,amssymb,showpacs]{revtex4-1}
\usepackage[latin9]{inputenc}
\usepackage{color}
\usepackage{amsmath}
\usepackage{amssymb}
\usepackage{graphicx}
\usepackage{esint}
\usepackage[unicode=true,pdfusetitle,
 bookmarks=true,bookmarksnumbered=false,bookmarksopen=false,
 breaklinks=false,pdfborder={0 0 0},pdfborderstyle={},backref=false,colorlinks=true]
 {hyperref}
\hypersetup{
 colorlinks,linkcolor=red,citecolor=blue}
\usepackage{breakurl}

\makeatletter
\usepackage{epsfig}\usepackage{amsfonts}

\makeatother

\begin{document}

\title{Doubly-resonant-cavity electromagnetically induced transparency}

\author{Xin-Xin Hu$^{1,2}$ }

\author{Chang-Ling Zhao$^{1,2}$ }

\author{Zhu-Bo Wang$^{1,2}$ }

\author{Yan-Lei Zhang$^{1,2}$ }

\author{Xu-Bo Zou$^{1,2}$ }

\author{Chun-Hua Dong$^{1,2}$ }

\author{Hong X. Tang$^{3}$ }

\author{Guang-Can Guo$^{1,2}$ }

\author{Chang-Ling Zou$^{1,2}$ }
\email{clzou321@ustc.edu.cn}

\affiliation{$^{1}$ Key Laboratory of Quantum Information, University of Science
and Technology of China, Hefei, 230026, People's Republic of China; }

\affiliation{$^{2}$ Synergetic Innovation Center of Quantum Information \& Quantum
Physics, University of Science and Technology of China, Hefei, Anhui
230026, China}

\affiliation{$^{3}$ Department of Electrical Engineering, Yale University, New
Haven, Connecticut 06511, USA}

\date{\today}
\begin{abstract}
We present an experimental study on the cavity-atom ensemble system,
and realize the doubly-resonant cavity enhanced electromagnetically
induced transparency, where both the probe and control lasers are
resonant with a Fabry-Perot cavity. We demonstrate the precise frequency
manipulating of the hybrid optical-atomic resonances, through either
temperature or cavity length tuning. In such a system, the control
power can be greatly enhanced due to the cavity, and all-optical switching
is achieved with a much lower control laser power compared to previous
studies. A new theoretical model is developed to describe the effective
three-wave mixing process between spin-wave and optical modes. Interesting
non-Hermitian physics are predicted theoretically and demonstrated
experimentally. Such a doubly-resonant cavity-atom ensemble system
without a specially designed cavity can be used for future applications,
such as optical signal storage and microwave-to-optical frequency
conversion.
\end{abstract}
\maketitle

\section{Introduction}

Near-resonance coherent light-atom interaction has been extensively
studied \cite{Lukin2003,Hammerer2010}, because the atomic media promises
the strong nonlinearity that cannot be achievable in dielectric materials.
The strong nonlinearity is potential for the low intensity light-light
interaction, which can be used for all-optical devices \cite{Bajcsy2009,Wu2010,Peyronel2012,Venkataraman2013}.
Usually, there are two approaches to enhance the coherent light-atom
interaction, one is using the optical cavity to enable the light passing
the atomic medium many times \cite{Thompson1992,Birnbaum2005,Zhang2011,Hacker2016},
and the other method is introducing atomic ensemble to collectively
enhance the interaction strength \cite{Raizen1989,Zhu1990,MinXiao2003,Tanji-Suzuki2011,Jing2014,Simon2007}.

In past two decades, great progresses have been achieved in such cavity-atomic
ensemble system. Various nonlinear optics effects have been studied
experimentally or theoretically, including the intracavity electromagnetically
induced transparency (EIT) \cite{Wu2008,Hernandez2009}, multiple
laser thresholds \cite{Fang-Yen2006}, the photon blockade effect
\cite{Guerlin2010}, strong and super-strong coupling \cite{Yu2009},
four-wave mixing \cite{Yu2010}, bistability \cite{Wang2002,Sawant2016}
and multi-stability \cite{Joshi2003,Wang2002}, as well as all-optical
switching \cite{Wang2002,Dawes2005,Wei2010,Dutta2017,Wang2017}. However,
in all previous experimental realizations of the all-optical switching,
at most one cavity resonance for the signal light is used, while the
control light is applied through free-space laser beam \cite{Wu2008,Wei2010,Sawant2016,Dutta2017}.
Although the cavity resonance for the control light has the advantages
that save the control laser power and are easy for control-signal
alignment, there is an experimental challenge for realizing the cavity
that doubly-resonant with two atomic transitions (e.g. $6.835\,\mathrm{GHz}$
for the $D_{2}$ line of $^{87}\mathrm{Rb}$). Because the frequency
difference between two optical resonances is determined by the cavity
geometry, the doubly-resonant condition can only be satisfied with
a specific cavity size.

In this work, we experimentally demonstrate the doubly-resonant Fabry-Perot
(FP) cavity-atom ensemble system without specially designed cavity
geometry, and achieve the all-optical switching with lower control
power. We develop a new theoretical treatment, in which the resonances
are actually the hybridized optical-atomic modes, to describe the
effective three-wave mixing process for multiple-level atom couple
with multiple cavity modes. Our theoretical analyses predict that
the resonance frequency is tunable due to the dispersive effect of
the atomic transitions and the nonlinearity of the $\Lambda$-type
atom leads to non-Hermitian coupling between hybridized modes and
spin-wave excitations. Benefiting from the different frequency shift
rate for two hybridized modes by changing the cavity length , we experimentally
demonstrated that the doubly-resonant condition can always be fulfilled by
controlling of atom density and the effective detuning between the
bare cavity resonances and atomic transitions. Based on this, we demonstrate
the all-optical switching by the control laser, and study the extinction
and response time of the switching. The demonstrated doubly-resonant
cavity-atom ensemble system can boost the nonlinear interaction between
light and spin excitation in the atom ensemble, can be used for other
applications such as optical signal storage and readout \cite{Simon2007,Hosseini2011},
and can also be generalized to other cavity systems and atom species.

\section{The system and principle}

\begin{figure}
\includegraphics[width=1\columnwidth]{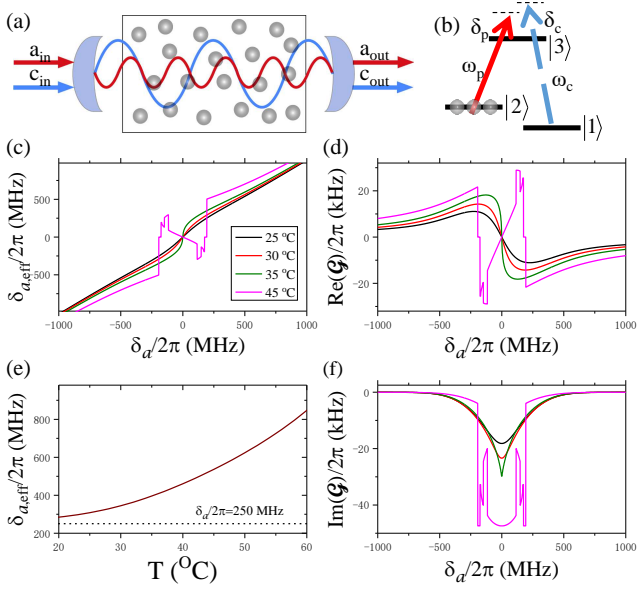}

\caption{(Color online) (a) Schematic illustration of the experimental system,
consisting of an ensemble of $^{87}\mathrm{Rb}$ that couples with
two resonant modes in a Fabry-Perot cavity. (b) The energy diagram
of the atomic system. Here, $\left|1\right\rangle $ and $\left|2\right\rangle $are
the two ground hyperfine states and $\left|3\right\rangle $ is the
excited state of $^{87}\mathrm{Rb}$ $D_{2}$-transitions. $\omega_{p}$
and $\omega_{c}$ are the frequency of probe and control lasers with
detunings $\delta_{p}$ and $\delta_{c}$, respectively. (c) and (e)
The dependence of $\delta_{a}^{\mathrm{eff}}$ on the $\delta_{a}$
for given temperatures and on the temperature for a given $\delta_{a}/2\pi=250\,\mathrm{MHz}$.
(d) and (f) The real and imaginary parts of $\mathcal{G}$. The non-negligible
imaginary part indicates the non-Hermitian interaction between the
three modes.}

\label{Fig1}
\end{figure}

The scheme for the system is depicted in Fig.$\,$1(a), in which a
three-level atom ensemble is confined in an FP cavity and interacts
with the cavity mode fields. If the transitions between ground hyperfine
states and excited states (as shown in Fig.$\,$\ref{Fig1}(b)) are
near-resonant with cavity modes, the interaction between the external
input optical field and the atoms can be greatly enhanced.  The Hamiltonian
for the cavity-atom ensemble can be written as ($\hbar=1$)
\begin{align}
H_{0} & =\omega_{a}a^{\dagger}a+\omega_{c}c^{\dagger}c\nonumber \\
 & +\sum_{j}\left[\left(\omega_{3}+kv_{j}\right)\sigma_{33,j}+\omega_{2}\sigma_{22,j}\right]\nonumber \\
 & +\sum_{j}\left[g_{a,j}a\sigma_{32,j}+g_{c,j}c\sigma_{31,j}+h.c.\right]\label{eq:Original-Hamiltonian}
\end{align}
Here, $\sigma_{mn,j}=\left|m\right\rangle _{j}\left\langle n\right|$
with subscript $j$ standing for $j$-th atom and $\left|1,2,3\right\rangle _{j}$
is the state of the atom as shown in Fig.$\,$1(b). $a$ and $c$
are the Bosonic operators for probe and control cavity modes, respectively,
with that the frequencies are $\omega_{a}$ and $\omega_{c}$, respectively.
$\omega_{3,j}=\omega_{3}+kv_{j}$ is the frequency with Doppler shift
for velocity of $v_{j}$, $k=\omega_{3}/v_{c}$ is the wave-vector
and $v_{c}$ is the velocity of light. $g_{a,j}\left(g_{c,j}\right)$
is the coupling strength between the atom transition and the cavity
mode for the $j$-th atom.

From the Eq.$\,$(\ref{eq:Original-Hamiltonian}), the three-level
atoms mediate the nonlinear interaction between the two optical modes
with distinct frequencies. The cascading interactions $a\sigma_{32}$
and $c\sigma_{31}$ would leads to an effective process of $ac^{\dagger}\sigma_{12}$,
i.e. the ground states of atoms have a flip when scattering the photons
and change their color. To directly describe such process, we develop
the effective three-wave mixing Hamiltonian in such a system. For
our experiments, we have $g_{a,j},g_{c,j}\ll\gamma_{23},\gamma_{13}$,
thus the excitation $\sigma_{33,j}\approx0$ can be neglected, then
we can solve the steady state of the system. Note that due to the
coupling between the cavity mode and atom transitions, the cavity
field is hybridized with the atomic excitation instead of the barely
electromagnetic field of the photon. Then, we can solve the effective
mode frequencies $\omega_{a}^{\mathrm{eff}}$ and $\omega_{c}^{\mathrm{eff}}$
for the hybridized cavity-atom ensemble modes, which satisfying
\begin{align*}
\sum_{j}\mathrm{Im}\left[\frac{g_{a,j}^{2}\sigma_{22,j}}{-i\left(\omega_{3,j}-\omega_{2}\right)-\gamma_{23}+i\omega_{a}^{\mathrm{eff}}}\right] & =\omega_{a}-\omega_{a}^{\mathrm{eff}},\\
\sum_{j}\mathrm{Im}\left[\frac{g_{c,j}^{2}\sigma_{11,j}}{-i\omega_{3,j}-\gamma_{13}+i\omega_{c}^{\mathrm{eff}}}\right] & =\omega_{c}-\omega_{c}^{\mathrm{eff}}.
\end{align*}
For hot atom ensemble, we have the sum over the ensemble corresponds
to the integration over the three-dimensional space and three-dimensional
momentum space, i. e. $\sum_{j}=\iiint\rho d^{3}\mathbf{x}\iiint P_{\mathrm{MB}}\left(\mathbf{v}\right)d^{3}\mathbf{v},$.
The Maxwell-Boltzman velocity distribution for $v_{z}$ is $P_{\mathrm{MB}}\left(v_{z}\right)=\sqrt{\frac{1}{2\pi\sigma_{v}}}\exp\left(-\frac{v_{z}^{2}}{2\sigma_{v}}\right)$,
with $\int_{-\infty}^{\infty}P_{\mathrm{MB}}\left(v_{z}\right)dv_{z}=1$
and $\sigma_{v}=\frac{k_{B}T}{m}$. Introducing the function
\begin{align}
F\left(\xi\right) & =\int dv_{z}\frac{P_{\mathrm{MB}}\left(v_{z}\right)}{-ikv_{z}+\xi}\nonumber \\
 & =\frac{1}{k}\sqrt{\frac{\pi}{2\sigma_{v}}}e^{\frac{\xi^{2}}{2k^{2}\sigma_{v}}}\left(1-\text{erf}\left(\frac{\xi}{\sqrt{2\sigma_{v}}k}\right)\right),
\end{align}
we have the characterization equations of the effective detunings
as
\begin{align}
\delta_{a}^{\mathrm{eff}}-\delta_{a}+\sigma_{22}G_{a}\mathrm{Im}\left[F\left(i\delta_{a}^{\mathrm{eff}}-\gamma_{23}\right)\right] & =0,\\
\delta_{c}^{\mathrm{eff}}-\delta_{c}+\sigma_{11}G_{c}\mathrm{Im}\left[F\left(i\delta_{c}^{\mathrm{eff}}-\gamma_{13}\right)\right] & =0,
\end{align}
where $\delta_{a}^{\mathrm{eff}}\left(\delta_{a}\right)=\omega_{a}^{\mathrm{eff}}\left(\omega_{a}\right)-\omega_{3}+\omega_{2}$
and $\delta_{c}^{\mathrm{eff}}\left(\delta_{c}\right)=\omega_{c}^{\mathrm{eff}}\left(\omega_{c}\right)-\omega_{3}$
are the effective detuning between the hybrid cavity mode (original
detuning between the bare cavity mode) and atomic transitions, and
$G_{a,c}=\rho\iiint g_{a,c}^{2}\left(\mathbf{x}\right)d^{3}\mathbf{x}$.
Introducing the collective spin wave operator
\begin{align}
m & =\sqrt{\rho}\frac{\iiint g_{a}\left(\mathbf{x}\right)g_{c}\left(\mathbf{x}\right)\sigma_{12}\left(\mathbf{x}\right)d^{3}\mathbf{x}}{\sqrt{\iiint g_{a}^{2}\left(\mathbf{x}\right)g_{c}^{2}\left(\mathbf{x}\right)d^{3}\mathbf{x}}},
\end{align}
we obtain the effective Hamiltonian of the system
\begin{eqnarray}
H_{\mathrm{eff}} & = & \omega_{a}^{\mathrm{eff}}a^{\dagger}a+\omega_{c}^{\mathrm{eff}}c^{\dagger}c+\omega_{2}m^{\dagger}m\nonumber \\
 &  & +\mathcal{G}\left(a^{\dagger}cm^{\dagger}+ac^{\dagger}m\right),
\end{eqnarray}
with the effective three-wave mixing coupling strength
\begin{equation}
\mathcal{G}=\sqrt{\rho\iiint g_{a}^{2}\left(\mathbf{x}\right)g_{c}^{2}\left(\mathbf{x}\right)d^{3}\mathbf{x}}\times F\left(i\delta_{a}^{\mathrm{eff}}-\gamma_{23}\right).\label{eq:G}
\end{equation}
The expression indicates that coupling strength $\mathcal{G}$ would
be a complex number, which corresponding to a non-Hermitian interaction
between light and spin wave excitations \cite{El-Ganainy2018}. Here,
the $\mathrm{Re}\left(\mathcal{G}\right)$ is the coherent coupling
strength originating from the coherent two-photon transition, while
the $\mathrm{Im}\left(\mathcal{G}\right)$ is the incoherent coupling
strength due to the spontaneous emission of excited state $\left|3\right\rangle $.
In the following, the effects due to the interesting non-Hermitian
mechanism would be revealed experimentally.

\begin{figure*}[!t]
\includegraphics[width=1.6\columnwidth]{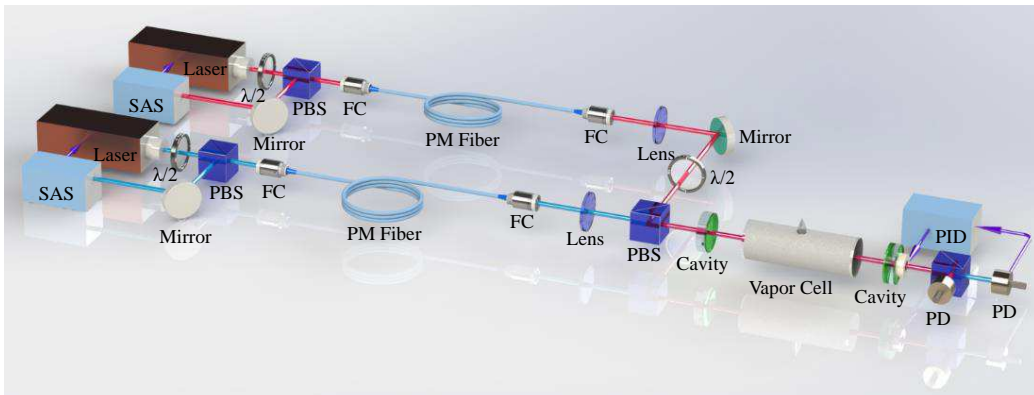}

\caption{(Color online) The experimental setup. Two external cavity diode lasers
provide the probe and control lasers, which are locked by saturated
absorption spectrum (SAS) and coupled into polarization-maintaining
(PM) fibers. The FP cavity with the length of $18\,\mathrm{cm}$ composes
of two concave mirrors with the same curvature radius of $100\,\mathrm{mm}$
and reflectivity of $97\%$ at $780\,\mathrm{nm}$. The cavity length
is stabilized by the piezoelectric transducer and PID feedback loop.
A glass vapor cell filled with pure $^{87}\mathrm{Rb}$ is placed
inside the cavity and covered with a soft heater (not shown here).
Here, $\lambda/2$, PBS, FC, and PD means the half-wave plate, polarization
beam splitter, fiber coupler, fiber and photon detector, respectively. }

\label{Fig2}
\end{figure*}
\begin{figure*}
\includegraphics[width=1.6\columnwidth]{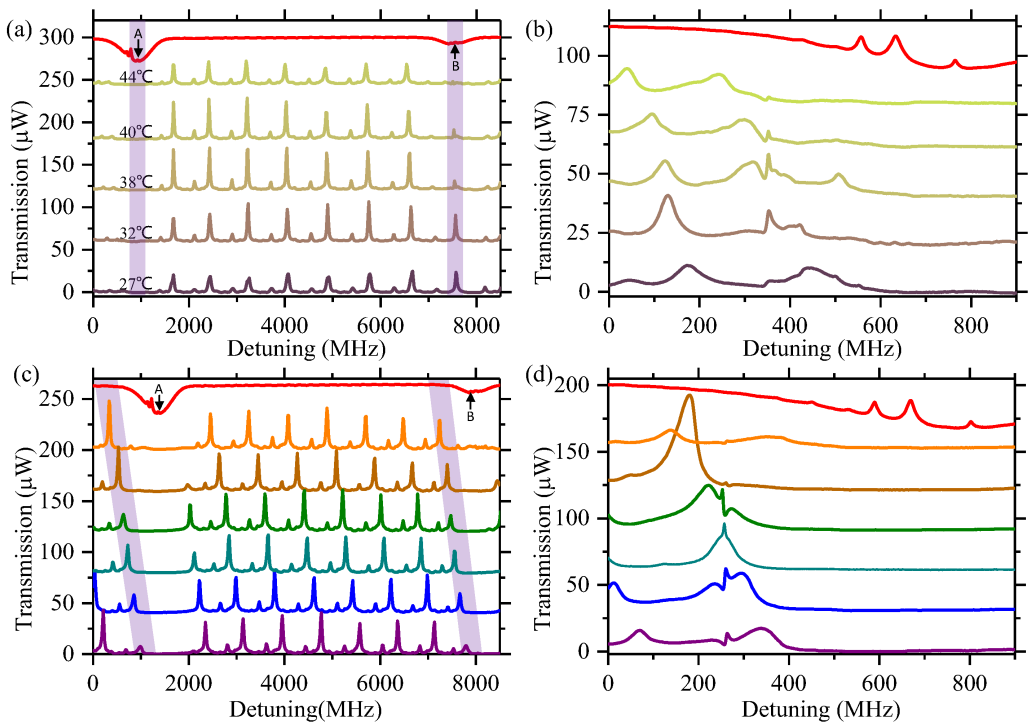}

\caption{(Color online) (a) The transmission of the cavity for different temperature
with the control cavity mode and lasers are locked to the transition
$5S_{1/2},F=1\rightarrow5P_{3/2},\,F^{'}=1$. The red solid line is
the saturated absorption spectrum for laser frequency calibration
and others show the transmission of the cavity at the different temperature.
The labels A and B denotes the two transitions $5S_{1/2},\,F=2\rightarrow5P_{3/2},\,F^{'}=1$
and $5S_{1/2},\,F=1\rightarrow5P_{3/2},\,F^{'}=1$, respectively.
(b) is the detailed spectra of (a), showing the sharp modification
of the spectrum due to the control field. (c) The transmission of
cavity for different cavity length, with the temperature, stabilized
at $\mathrm{40\,\ensuremath{^{\circ}}C}$. The effective signal mode
detuning can be tuned by changing the cavity length, with the assistance
of nonlinear dispersion by the cavity-atom ensemble hybridization.
For appropriate cavity length, there are two modes with a frequency
shift matching 6.835$\mathrm{GHz}$ and satisfying the doubly-resonant
condition. The blue shade shows that the two modes have different
frequency shifts at different cavity length. (d) is the detailed spectrum
of (c) and an obvious EIT-lineshape can be observed. In this figure,
spectra are plotted with equidistant bias for convenience.}

\label{Fig3}
\end{figure*}

In our experimental configuration, there is a strong control field
on mode $c$, thus approximately we have $\sigma_{11}\approx0$, then
$\omega_{c}^{\mathrm{eff}}=\omega_{c}$. Including the control and
probe laser fields, the system effective Hamiltonian in the rotating
frame is
\begin{align}
H & =\left(\delta_{a}^{\mathrm{eff}}-\delta_{p}\right)a^{\dagger}a+\left(\delta_{c}-\delta_{l}\right)c^{\dagger}c\nonumber \\
 & +\left(\omega_{2}+\delta_{p}-\delta_{l}\right)m^{\dagger}m+\mathcal{G}\left(a^{\dagger}cm^{\dagger}+ac^{\dagger}m\right)\nonumber \\
 & +\left(\sqrt{2\kappa_{a,1}}a_{in}a^{\dagger}+\sqrt{2\kappa_{c,1}}c_{in}c^{\dagger}+h.c.\right),
\end{align}
with input laser detunings $\delta_{p}=\omega_{p}-\omega_{3}$ and
$\delta_{l}=\omega_{l}-\omega_{3}+\omega_{2}$, field amplitudes $a_{in}=\sqrt{P_{p}/\hbar\omega_{p}}$,
$c_{in}=\sqrt{P_{c}/\hbar\omega_{l}}$, powers $P_{p}$ and $P_{c}$,
respectively. The external coupling rate to the cavity modes through
the input mirror is $\kappa_{a(c),1}/2\pi=\frac{v_{c}}{4L}\frac{-\ln r}{2\pi}$,
$L$ is the length of the cavity and $r$ is the reflectivity of the
input mirror of the FP cavity (the reflectivity for modes $a$ and
$c$ are almost the same). The steady state solution of the probe
light transmittance $\left(\frac{d}{dt}a=\frac{d}{dt}c=0\right)$
is
\begin{align}
t & =\frac{2\sqrt{\kappa_{a,1}\kappa_{a,2}}}{-i\left(\delta_{a}^{\mathrm{eff}}-\delta_{p}\right)-\kappa_{a}-\frac{n_{c}\mathcal{G}^{2}}{-i\left(\omega_{2}+\delta_{p}-\delta_{l}\right)+\kappa_{m}}}.\label{eq:Trans}
\end{align}
Here, $n_{c}=2\kappa_{c,1}c_{in}^{2}/\left[\left(\omega_{c}-\omega_{l}\right)^{2}+\kappa_{c}^{2}\right]$
is the intracavity control photon number, $\kappa_{a,2}$ is the coupling
rate for the output mirror, with $\kappa_{a}$, $\kappa_{c}$ and
$\kappa_{m}$ are the total loss rate of mode $a$, $c$ and the collective
spin wave excitation $m$, respectively.

From the equation, the transmission of the cavity mode can be modified
by the control field. When $\delta_{a}^{\mathrm{eff}}=\delta_{l}-\omega_{2}$
is satisfied, the modification of the probe field transmittance is
most significant, which is proportional to three-wave interaction
cooperativity $\mathcal{C}=n_{c}\mathcal{G}^{2}/\kappa_{a}\kappa_{m}$.
Therefore, there are additional two requirements for strong interaction:
(1) the resonant condition $\delta_{c}-\delta_{l}=0$ for control
laser, since the $n_{c}$ can be boosted by the on-resonant control
field; (2) larger atom density $\rho$ or higher temperature, since
$\mathcal{G}^{2}\propto\rho$.

The temperature dependent density of $^{87}\mathrm{Rb}$ atoms can
be estimated by $\rho\left(T\right)=10^{7.738-\frac{4215}{T}}/k_{B}T$
\cite{Siddons2008}, thus we can numerically solve the $\delta_{a}^{\mathrm{eff}}$
and $\mathcal{G}$ with experimental parameters (explained in the
next section). In Fig.$\,$\ref{Fig1}(c), the $\delta_{a}^{\mathrm{eff}}$
varies with the $\delta_{a}$ and shows nonlinear dependence when
$\delta_{a}\sim0$, as a result of the hybridization of the bare cavity
and the atom ensemble. For higher temperature, the difference $\delta_{a}^{\mathrm{eff}}-\delta_{a}$
becomes larger, and eventually shows strange shape due to the strong
nonlinearity in dense atom ensemble. In Fig.$\,$\ref{Fig1}(e), the
dependence of $\delta_{a}^{\mathrm{eff}}$ on the temperature for
given $\delta_{a}/2\pi=250\,\mathrm{MHz}$ is depicted. Both plots
show that the $\delta_{a}^{\mathrm{eff}}$ can be efficiently controlled
by the bare cavity frequency and the atom vapor cell temperature,
so there are two approaches to realize the doubly-resonant interaction.
In addition, we also estimated the $\mathcal{G}$ for different conditions
{[}Fig.$\,$\ref{Fig1}(d)\&(f){]}, showing that the interaction is
enhanced when the cavity is near-resonant with the atom and the temperature
is increased. However, the $\mathcal{G}$ shows non-negligible imaginary
part, indicating the non-Hermitian interaction between the three modes,
which would lead to new effect in such a system other than the traditional
lossless three-wave coupling system \cite{Weis2010,Singh2014,Zhang2014,Fan2015,Balram2016}.

\section{Experimental Setup}

Our experimental setup is shown in Fig.$\,$\ref{Fig2}. The standing-wave
FP cavity with a length of $18\,\mathrm{cm}$ ($\mathrm{FSR=854\,MHz}$)
composes of two concave mirrors with the same curvature radius of
$100\,\mathrm{mm}$ and reflectivity of $97\%$ at $780\,\mathrm{nm}$.
One mirror is mounted on a piezoelectric transducer (PZT) to adjust
the cavity length and can be locked into the control laser by a feedback
loop. A glass vapor cell (Photonics Technologies, Rubidium-87) with
a length of 75$\,\mathrm{mm}$, which is filled with pure $^{87}\mathrm{Rb}$
gas and wrapped by a soft heater, is placed inside the cavity. The
heater is controlled by a temperature controller (Thorlabs, TC200)
to stabilize the temperature of the cell with the uncertainty less
than $0.1\,\mathrm{\ensuremath{^{\circ}}C}$. Two external cavity
diode lasers (Toptica DL100) are employed in the experiment, one for
probe and the other one for control. The probe laser is coupled to
the cavity through a standard polarization maintaining single-mode
fiber (OZOptics), with the frequency continuously scanning cross the
transitions between $5S_{1/2}$, $F=2$ and $5P_{3/2}$, $F'=1,2,3$
of $^{87}\mathrm{Rb}$. Similarly, the control laser is also coupled
to the cavity, with the frequency locked around the transition from
$5S_{1/2}$, $F=1$ to $5S_{3/2}$,$\,F^{'}=1$ of $^{87}\mathrm{Rb}$
by saturated absorption spectrum (SAS). The two lasers of orthogonal
polarization are combined by a polarization beam splitter (PBS), and
coupled to the cavity mirror. The beam profiles of both probe and
control lasers are adjusted by an anti-reflection coated convex lens
(focal length $25\,\mathrm{mm}$) to match the fundamental transverse
electromagnetic mode ($\mathrm{TEM_{00}}$) of the optical cavity.
The transmitted probe and control lasers through the cavity are separated
by PBS and measured by two detectors (Thorlabs, PDA36A). In our experiment,
we tested the FP cavity by measuring the transmission spectrum of
the probe, and estimated the finesses of the cavity $\mathcal{F}=21$
with cell placed in the cavity ($\mathcal{F}=120$ without cell).

\begin{figure}
\includegraphics[width=1\columnwidth]{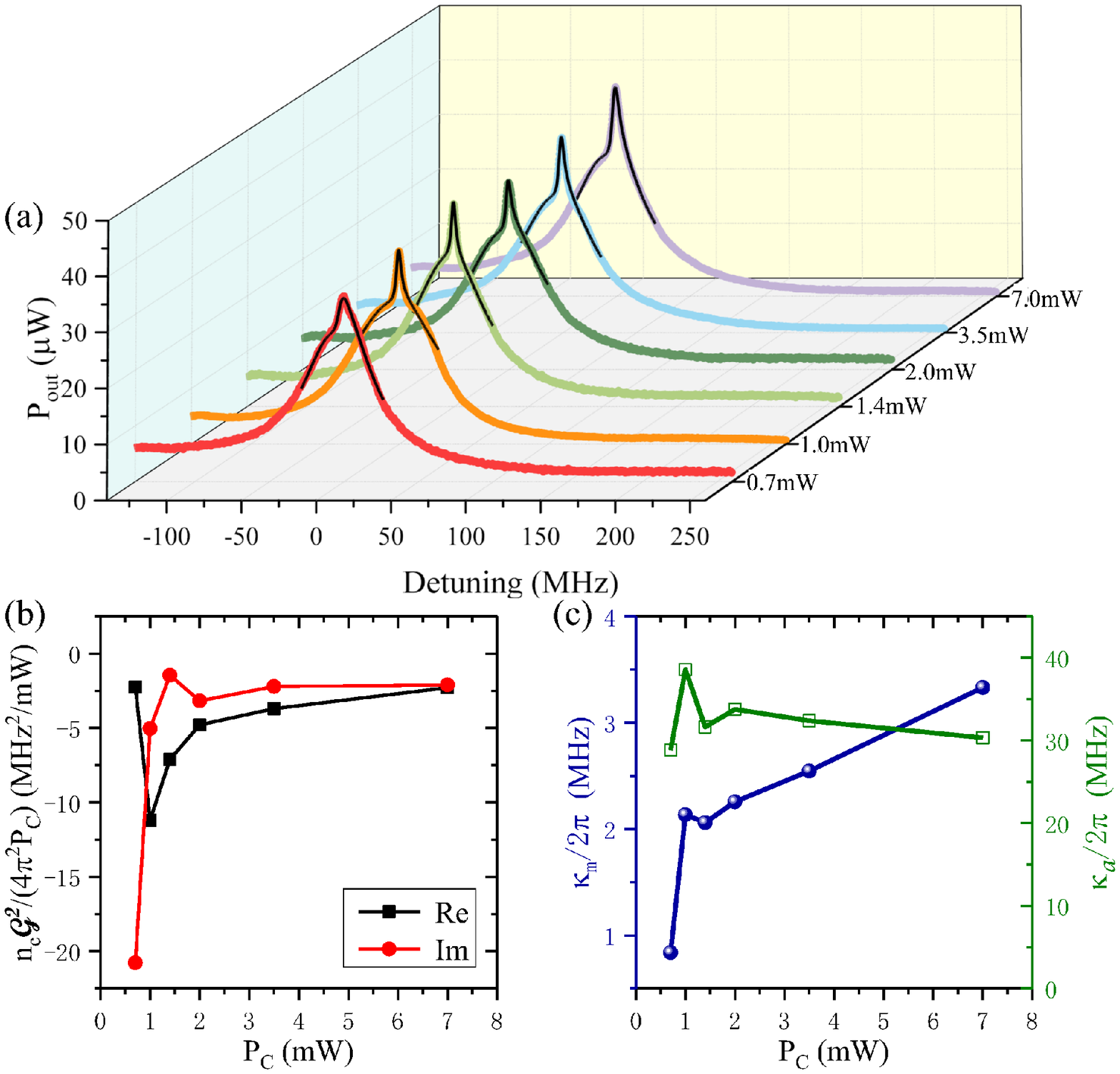}

\caption{(Color online) The transmission of the probe with different control
laser power. (a) Enlarged spectrum under doubly-resonant condition.
The spectra can be perfectly fitted by Eq.$\,$(\ref{eq:G}) (black
solid line). (b) The real (black) and imaginary (red) part of the
fitted parameter $n_{c}\mathcal{G}^{2}/P_{c}$ against the control
power, which effectively changes the intrinsic hybrid mode loss and
frequency. (c) The fitted $\kappa_{m}$ (blue line) and $\kappa_{a}$
(green line) against the control power.}

\label{Fig4}
\end{figure}

As discussed above, we should tune the cavity to find two modes with
frequency differences being exactly the $6.835\,\mathrm{GHz}$ for
the energy difference of ground hyperfine states. We have two strategies:
(1) change the atom density of the cavity by changing the temperature
of the vapor cell and (2) tune the cavity length.

In the first approach, we lock the control laser to the frequency
of the transition $5S_{1/2},\,F=1\rightarrow5P_{3/2},\,F^{'}=1$,
and then one of the cavity modes is locked to the control laser by
the feedback based on the transmission of the control laser. When
the temperature increases, the atom density in the cell increases,
thus the cavity modes close to the probe $2\rightarrow3$ transition
will strongly couple with the atoms, leading to the frequency shift,
as predicted in Fig.$\,$\ref{Fig1}(c) and (e). As shown in Fig.$\,$\ref{Fig3}(a)
and (b), by increasing the temperature, the frequency shifting increases.
At specific temperature, we can find that the frequency difference
between two modes be on-resonance with the $5S_{1/2},\,F=1\rightarrow5P_{3/2},\,F^{'}=1$.
Although the doubly-resonant cavity EIT has been observed {[}Fig.$\,$2(b){]},
the optical modes for probe and control are both with high-loss and
small transmittance. The reason is that the modes are too close to
the atom transition frequencies leading to a strong absorption loss.
From Fig.$\,$2(a), the control mode almost disappears when the temperature
approaches $40\,\mathrm{\ensuremath{^{\circ}}C}$, and we can not
lock the cavity by control mode above this temperature since the atom
still induces considerable loss to the control mode with the residue
population of the atom ensemble on $\left|2\right\rangle $.

The cavity spectrum for the other approach is shown in Fig.$\,$\ref{Fig3}(c),
where the control laser is locked with a frequency offset to the transition.
In this case, both probe and control modes experience the strong coupling
to the atom ensemble induced mode frequency shift for high atom density
(temperature at $40\,\mathrm{\ensuremath{^{\circ}}C}$). When changing
the control mode frequency offset, the cavity length changes accordingly
by the PZT. Benefiting from the hybridization with atom ensemble,
the frequency shift rates for two modes by changing the cavity length
are different, and we can always find a proper control frequency offset
that the doubly-resonant condition can be fulfilled. Shown in Fig.$\,$\ref{Fig3}(c)
are the experimental results that the frequency spacing between two
cavity modes can be tuned from $\mathrm{6.4\,GHz}$ to $7.0\,\mathrm{GHz}$.
The detailed EIT spectrum of the probe laser is shown in Fig.$\,$\ref{Fig3}(d).
Comparing to the first approach, due to the offset to the atom transitions,
the absorption of atom ensemble will not be too strong and both the
control and probe modes can maintain high-quality factor and large
transmittance.

\section{Doubly-resonant cavity EIT}

By the second approach, we locked the control mode to a specific frequency,
which has a detuning of about $6.835\,\mathrm{GHz}$ with respect
to the probe mode frequency, and realize the doubly-resonant condition
at $\mathrm{40\ensuremath{^{\circ}}C}$. To be more specific, the
control laser is locked around the transition B in Fig.$\,$\ref{Fig3}(c)
and the probe laser is scanning around the probe mode close the transition
A in Fig.$\,$\ref{Fig3}(c). In the following experiments, the effective
detuning for mode $a$ is about $500\,\mathrm{MHz}$ at about 40 degrees,
corresponds to the initial detuning of about $250\,\mathrm{MHz}$.
To verify the doubly-resonant cavity EIT, we measure the spectrum
of the probe mode for various control laser powers, as shown in Fig.$\,$\ref{Fig4}(a).
As expected, there is a modification to the Lorentzian shape on the
probe mode resonance when the probe laser frequency detuning to the
control laser frequency matches the $\omega_{2}/2\pi\sim6.835\,\mathrm{GHz}$.
This can be explained that the reduced absorption by the control laser.
According to our theory, the transmission of probe light can be estimated
by Eq.$\,$(\ref{eq:Trans}). From the equation, the control laser
induces an extra term $\frac{n_{c}\mathcal{G}^{2}}{-i\left(\omega_{2}+\delta_{p}-\delta_{l}\right)+\kappa_{m}}$,
so effectively changing the intrinsic hybrid mode loss and frequency.
For the on-resonance case $\delta_{a}^{\mathrm{eff}}-\delta_{p}=0$
and $\omega_{2}+\delta_{p}-\delta_{l}=0$, the maximum modification
of the transmission (ratio between case with and without control)
is
\begin{equation}
\eta=\left|\frac{1}{1+\mathcal{C}}\right|^{2}=\left|\frac{1}{1+\frac{n_{c}\mathcal{G}^{2}}{\kappa_{m}\kappa_{a}}}\right|^{2}.
\end{equation}
According to the traditional EIT without cavity \cite{Olson2009},
the control laser would reduce the absorption of the atomic medium,
thus there is a control laser-induced transmission peak. However,
for our system, the hybrid mode containing both atomic medium and
photon with coherent feedback by the mirrors, the behaviors are essentially
different from the cavity-less case. As indicated by Eq.$\,$(\ref{eq:G}),
due to the spontaneous emission of the excited level $\left|e\right\rangle $,
the effective coupling strength $\mathcal{G}$ is a complex number,
which leads to the non-Hermitian coupling between the collective spin
wave mode and the hybrid optical modes. If $\mathcal{\mathcal{G}}$
is pure coherent coupling, i.e. is a real number, then $\mathcal{C}>0$
and the control laser will induce a dip in the transmission spectrum.
If $\mathcal{G}$ is pure incoherent coupling that hybrid mode frequency
detuning is close to $0$ (Fig.$\,$\ref{Fig1}(f)), i.e. is an imaginary
number, then $\mathcal{C}<0$ and there will be a peak in the transmission
spectrum.

The results in Fig.$\,$\ref{Fig4}(a) indicates that the incoherent
coupling dominates, as there is always a peak induced by the control.
To verify the model of our doubly-resonant system, we fitted the spectra.
We found that the parameter $n_{c}\mathcal{G}^{2}/P_{c}$ is a complex
number, and shows relatively large fluctuation at low pump strength
while approaches a steady state value when pump strength becomes larger.
The fitted cavity linewidth is about $34\,\mathrm{MHz}$, while the
$\kappa_{m}$ linearly increases with the control laser power, as
shown in Fig.$\,$\ref{Fig4}(c). This is consistent with our model
that the control laser will induce an incoherent decay channel from
the ground state $\left|1\right\rangle $ to $\left|2\right\rangle $.
For the large fluctuation of parameters at low pump power, it can
be attributed to the mechanism that the assumption of steady state
population on level $\left|2\right\rangle $ is not valid for weak
pump power.

\section{All-optical Switching}

\begin{figure}
\includegraphics[width=1\columnwidth]{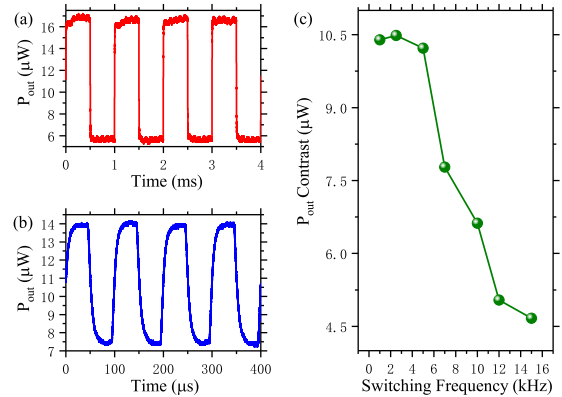}

\caption{(Color online) All-optical switching based on the doubly-resonant
cavity EIT. (a) and (b) The temporal response of cavity signal transmission
for square-wave control power modulation, with the modulation rate
of $1\,\mathrm{kHz}$ and $10\,\mathrm{kHz}$, respectively. (c) The
contrast of the signal transmittance for different switching rate. }

\label{Fig5}
\end{figure}
Based on the doubly-resonant cavity EIT, we can control the probe
laser transmittance optically, which can be applied to realize the
all-optical switching \cite{Wang2002,Dawes2005,Wei2010,Sawant2016,Dutta2017}.
Here, the probe laser is locked to the cavity resonance around the
transition $5S_{1/2},\,F=2\rightarrow5P_{3/2},\,F^{'}=1$, while the
control lasers is locked to the transition $5S_{1/2},\,F=1\rightarrow5P_{3/2},\,F^{'}=1$.
The control laser is actually generated by a laser passing an AOM,
and the control laser power and frequency are controlled by the RF
driving on the AOM. So, the modulation of control laser is realized
by modulating the RF power, with the RF frequency be adjusted to make
the frequency difference between the control and the probe lasers
matching the frequency difference between $5S_{1/2},\,F=1$ and $5S_{1/2},\,F=2$.

To characterize such a switch, we test the response of the transmitted
probe laser for a square-wave control power modulation. During the
experiments, the vapor cell is stabilized at a fixed temperature ($40\,\mathrm{\ensuremath{^{\circ}}C}$)
and the input control power is fixed at $1.4\,\mathrm{mW}$. The results
as shown in Fig.$\,$\ref{Fig5}(a) and (b) are the signal transmission
with the control laser power modulation frequency of $1\,\mathrm{kHz}$
and $10\,\mathrm{kHz}$, respectively. At low modulation rate, the
signal transmission follows the control modulation, showing nice square-wave
shape {[}Fig.$\,$\ref{Fig5}(a){]}. When the modulation rate is as
large as $10\,\mathrm{kHz}$, the signal transmission waveform deforms
from the square-wave shape obviously {[}Fig.$\,$\ref{Fig5}(b){]}.
Figure$\,$\ref{Fig5}(c) depicts the contrast of signal transmittance
against the switching rate. The all-optical switching extinction decreases
when the modulation rate exceeds $8\,\mathrm{kHz}$.

\section{Discussion}

Due to the impedance mismatching between the two cavity mirrors and
the intracavity losses, most of the input probe and control laser
power is reflected by the mirror. From the experimental results, the
typical hybrid mode decay rate is $\kappa_{a}\sim2\pi\times34\,\mathrm{MHz}$,
and the $97\%$ reflectivity of the mirror leads to $\kappa_{a,1}=\kappa_{a,2}=2\pi\times2.0\,\mathrm{MHz}$,
which means the reflection of input on-resonance probe or control
laser is $R\approx\left|\frac{\kappa_{a}-2\kappa_{a,1}}{\kappa_{a}}\right|^{2}\sim78\%$.
This indicates it is possible to design an all-optical switching with
weaker control power by choosing proper cavity mirror reflectivity
to satisfy the impedance matching condition. The on-off ratio of the
switch $\eta$ could be improved for higher $\mathcal{C}$, but it
is limited by the saturated cooperativity $\mathcal{C}=\frac{n_{c}\mathcal{G}^{2}}{\kappa_{m}\kappa_{a}}$
because $\kappa_{m}$ also increases with $n_{c}$. As a result, the
EIT peak height saturated for large pump power in Fig.$\,$\ref{Fig4}(a)
(the estimation of saturated $\eta\sim1.37$ for our experimental
results). In addition, the rate of the switching is limited by the
atom flying effect. After turning on the laser, the population of
the atom ensemble reaches the steady state after all the atoms passing
through the control laser. The vapor cell we used currently has an
average flying time of about $42\,\mathrm{\mu s}$, corresponding
to a relaxation rate of the order of $3.8\,\mathrm{kHz}$, which agrees
with the curve shown in Fig.$\,$\ref{Fig5}(c).

The possible approaches to improve the performance of the switching
include optimizing the cavity to reduce the intrinsic loss of the
hybrid mode due to the scattering and absorption of vapor cell wall,
and optimizing the $\delta_{a}$ to minimize the control light inducing
collective spin wave relaxation. Besides, we do not shield the magnetic
field induced by earth or spin relaxation due to the collision to
the cell wall. If we add magnetic shields around $^{87}\mathrm{Rb}$
vapor cell and also anti-relaxation coating in the cell, the linewidth
of the EIT will decrease significantly and the extinction will be
greatly enhanced.

The doubly-resonant condition can actually enable the cavity-enhanced
photon-spin wave coupling, thus we expect this method be used for
quantum memory with reduced control power. This doubly-resonant cavity
EIT switch may also be generalized to a more compact platform, such
as integrated photonic circuits \cite{Ritter2016,Stern2016,Stern2017}.
For smaller cavity, the FSR is much larger than the ground state splitting,
but we can explore the modes of different polarization to satisfy
the doubly-resonant condition. In addition, adding an external bias
magnetic field will lift the degeneracy of the transitions and adjust
the split between two modes. Therefore the doubly-resonant cavity-atom
ensemble coupling can be achieved in the microcavity easily.

\section{Conclusion}

In summary, we theoretically and experimentally studied the electromagnetically
induced transparency of a cavity-atom ensemble system, in which the
transitions of two ground states of $^{87}\mathrm{Rb}$ $D_{2}$ line
are on-resonant with two fundamental modes of the cavity. We demonstrate
two approaches that can precisely tune the hybrid cavity mode frequencies
to achieve the doubly-resonant condition. We measure and analyze the
effect of all-optical switching based on the doubly-resonant cavity
EIT, and realized a switching extinction of about $46\%$ at a rate
of $10\,\mathrm{kHz}$. The developed theoretical treatment of the
three-wave mixing between spin-wave and optical modes would find application
in future microwave-to-optical frequency conversion, and the revealed
incoherent coupling may stimulate further studies on non-Hermitian
physics by exploring such as cavity-atom ensemble system.
\begin{acknowledgments}
We thank Pengfei Zhang, Xudong Yu, Tian Xia, Yanhong Xiao, Chuan-Sheng
Yang for fruitful discussions. This work was supported by the National
Key R \& D Program (Grant No.2016YFA0301300, 2017YFA0304504 and 2016YFA0301700),
National Natural Science Foundation of China (Grant No. 61505195 and
91536219), and Anhui Initiative in Quantum Information Technologies
(AHY130000).
\end{acknowledgments}

\bibliographystyle{Zou}

\begin{thebibliography}{41}%
\makeatletter
\providecommand \@ifxundefined [1]{%
 \@ifx{#1\undefined}
}%
\providecommand \@ifnum [1]{%
 \ifnum #1\expandafter \@firstoftwo
 \else \expandafter \@secondoftwo
 \fi
}%
\providecommand \@ifx [1]{%
 \ifx #1\expandafter \@firstoftwo
 \else \expandafter \@secondoftwo
 \fi
}%
\providecommand \natexlab [1]{#1}%
\providecommand \enquote  [1]{``#1''}%
\providecommand \bibnamefont  [1]{#1}%
\providecommand \bibfnamefont [1]{#1}%
\providecommand \citenamefont [1]{#1}%
\providecommand \href@noop [0]{\@secondoftwo}%
\providecommand \href [0]{\begingroup \@sanitize@url \@href}%
\providecommand \@href[1]{\@@startlink{#1}\@@href}%
\providecommand \@@href[1]{\endgroup#1\@@endlink}%
\providecommand \@sanitize@url [0]{\catcode `\\12\catcode `\$12\catcode
  `\&12\catcode `\#12\catcode `\^12\catcode `\_12\catcode `\%12\relax}%
\providecommand \@@startlink[1]{}%
\providecommand \@@endlink[0]{}%
\providecommand \url  [0]{\begingroup\@sanitize@url \@url }%
\providecommand \@url [1]{\endgroup\@href {#1}{\urlprefix }}%
\providecommand \urlprefix  [0]{URL }%
\providecommand \Eprint [0]{\href }%
\providecommand \doibase [0]{http://dx.doi.org/}%
\providecommand \selectlanguage [0]{\@gobble}%
\providecommand \bibinfo  [0]{\@secondoftwo}%
\providecommand \bibfield  [0]{\@secondoftwo}%
\providecommand \translation [1]{[#1]}%
\providecommand \BibitemOpen [0]{}%
\providecommand \bibitemStop [0]{}%
\providecommand \bibitemNoStop [0]{.\EOS\space}%
\providecommand \EOS [0]{\spacefactor3000\relax}%
\providecommand \BibitemShut  [1]{\csname bibitem#1\endcsname}%
\let\auto@bib@innerbib\@empty
%</preamble>
\bibitem [{\citenamefont {Lukin}(2003)}]{Lukin2003}%
  \BibitemOpen
  \bibfield  {author} {\bibinfo {author} {\bibfnamefont {M.}~\bibnamefont
  {Lukin}},\ }\bibfield  {title} {\enquote {\bibinfo {title} {{Colloquium:
  Trapping and manipulating photon states in atomic ensembles}},}\ }\href
  {\doibase 10.1103/RevModPhys.75.457} {\bibfield  {journal} {\bibinfo
  {journal} {Rev. Mod. Phys.}\ }\textbf {\bibinfo {volume} {75}},\ \bibinfo
  {pages} {457} (\bibinfo {year} {2003})}\BibitemShut {NoStop}%
\bibitem [{\citenamefont {Hammerer}\ \emph {et~al.}(2010)\citenamefont
  {Hammerer}, \citenamefont {S{\o}rensen},\ and\ \citenamefont
  {Polzik}}]{Hammerer2010}%
  \BibitemOpen
  \bibfield  {author} {\bibinfo {author} {\bibfnamefont {K.}~\bibnamefont
  {Hammerer}}, \bibinfo {author} {\bibfnamefont {A.~S.}\ \bibnamefont
  {S{\o}rensen}}, \ and\ \bibinfo {author} {\bibfnamefont {E.~S.}\ \bibnamefont
  {Polzik}},\ }\bibfield  {title} {\enquote {\bibinfo {title} {{Quantum
  interface between light and atomic ensembles}},}\ }\href {\doibase
  10.1103/RevModPhys.82.1041} {\bibfield  {journal} {\bibinfo  {journal} {Rev.
  Mod. Phys.}\ }\textbf {\bibinfo {volume} {82}},\ \bibinfo {pages} {1041}
  (\bibinfo {year} {2010})}\BibitemShut {NoStop}%
\bibitem [{\citenamefont {Bajcsy}\ \emph {et~al.}(2009)\citenamefont {Bajcsy},
  \citenamefont {Hofferberth}, \citenamefont {Balic}, \citenamefont {Peyronel},
  \citenamefont {Hafezi}, \citenamefont {Zibrov}, \citenamefont {Vuletic},\
  and\ \citenamefont {Lukin}}]{Bajcsy2009}%
  \BibitemOpen
  \bibfield  {author} {\bibinfo {author} {\bibfnamefont {M.}~\bibnamefont
  {Bajcsy}}, \bibinfo {author} {\bibfnamefont {S.}~\bibnamefont {Hofferberth}},
  \bibinfo {author} {\bibfnamefont {V.}~\bibnamefont {Balic}}, \bibinfo
  {author} {\bibfnamefont {T.}~\bibnamefont {Peyronel}}, \bibinfo {author}
  {\bibfnamefont {M.}~\bibnamefont {Hafezi}}, \bibinfo {author} {\bibfnamefont
  {A.~S.}\ \bibnamefont {Zibrov}}, \bibinfo {author} {\bibfnamefont
  {V.}~\bibnamefont {Vuletic}}, \ and\ \bibinfo {author} {\bibfnamefont
  {M.~D.}\ \bibnamefont {Lukin}},\ }\bibfield  {title} {\enquote {\bibinfo
  {title} {{Efficient All-Optical Switching Using Slow Light within a Hollow
  Fiber}},}\ }\href {\doibase 10.1103/PhysRevLett.102.203902} {\bibfield
  {journal} {\bibinfo  {journal} {Phys. Rev. Lett.}\ }\textbf {\bibinfo
  {volume} {102}},\ \bibinfo {pages} {203902} (\bibinfo {year}
  {2009})}\BibitemShut {NoStop}%
\bibitem [{\citenamefont {Wu}\ \emph {et~al.}(2010)\citenamefont {Wu},
  \citenamefont {Hulbert}, \citenamefont {Lunt}, \citenamefont {Hurd},
  \citenamefont {Hawkins},\ and\ \citenamefont {Schmidt}}]{Wu2010}%
  \BibitemOpen
  \bibfield  {author} {\bibinfo {author} {\bibfnamefont {B.}~\bibnamefont
  {Wu}}, \bibinfo {author} {\bibfnamefont {J.~F.}\ \bibnamefont {Hulbert}},
  \bibinfo {author} {\bibfnamefont {E.~J.}\ \bibnamefont {Lunt}}, \bibinfo
  {author} {\bibfnamefont {K.}~\bibnamefont {Hurd}}, \bibinfo {author}
  {\bibfnamefont {A.~R.}\ \bibnamefont {Hawkins}}, \ and\ \bibinfo {author}
  {\bibfnamefont {H.}~\bibnamefont {Schmidt}},\ }\bibfield  {title} {\enquote
  {\bibinfo {title} {{Slow light on a chip via atomic quantum state
  control}},}\ }\href {\doibase 10.1038/nphoton.2010.211} {\bibfield  {journal}
  {\bibinfo  {journal} {Nat. Photonics}\ }\textbf {\bibinfo {volume} {4}},\
  \bibinfo {pages} {776} (\bibinfo {year} {2010})}\BibitemShut {NoStop}%
\bibitem [{\citenamefont {Peyronel}\ \emph {et~al.}(2012)\citenamefont
  {Peyronel}, \citenamefont {Firstenberg}, \citenamefont {Liang}, \citenamefont
  {Hofferberth}, \citenamefont {Gorshkov}, \citenamefont {Pohl}, \citenamefont
  {Lukin},\ and\ \citenamefont {Vuleti{\'{c}}}}]{Peyronel2012}%
  \BibitemOpen
  \bibfield  {author} {\bibinfo {author} {\bibfnamefont {T.}~\bibnamefont
  {Peyronel}}, \bibinfo {author} {\bibfnamefont {O.}~\bibnamefont
  {Firstenberg}}, \bibinfo {author} {\bibfnamefont {Q.-Y.}\ \bibnamefont
  {Liang}}, \bibinfo {author} {\bibfnamefont {S.}~\bibnamefont {Hofferberth}},
  \bibinfo {author} {\bibfnamefont {A.~V.}\ \bibnamefont {Gorshkov}}, \bibinfo
  {author} {\bibfnamefont {T.}~\bibnamefont {Pohl}}, \bibinfo {author}
  {\bibfnamefont {M.~D.}\ \bibnamefont {Lukin}}, \ and\ \bibinfo {author}
  {\bibfnamefont {V.}~\bibnamefont {Vuleti{\'{c}}}},\ }\bibfield  {title}
  {\enquote {\bibinfo {title} {{Quantum nonlinear optics with single photons
  enabled by strongly interacting atoms.}}}\ }\href {\doibase
  10.1038/nature11361} {\bibfield  {journal} {\bibinfo  {journal} {Nature}\
  }\textbf {\bibinfo {volume} {488}},\ \bibinfo {pages} {57} (\bibinfo {year}
  {2012})}\BibitemShut {NoStop}%
\bibitem [{\citenamefont {Venkataraman}\ \emph {et~al.}(2013)\citenamefont
  {Venkataraman}, \citenamefont {Saha},\ and\ \citenamefont
  {Gaeta}}]{Venkataraman2013}%
  \BibitemOpen
  \bibfield  {author} {\bibinfo {author} {\bibfnamefont {V.}~\bibnamefont
  {Venkataraman}}, \bibinfo {author} {\bibfnamefont {K.}~\bibnamefont {Saha}},
  \ and\ \bibinfo {author} {\bibfnamefont {A.~L.}\ \bibnamefont {Gaeta}},\
  }\bibfield  {title} {\enquote {\bibinfo {title} {{Phase modulation at the
  few-photon level for weak-nonlinearity-based quantum computing}},}\ }\href
  {\doibase 10.1038/nphoton.2012.283} {\bibfield  {journal} {\bibinfo
  {journal} {Nat. Photonics}\ }\textbf {\bibinfo {volume} {7}},\ \bibinfo
  {pages} {138} (\bibinfo {year} {2013})}\BibitemShut {NoStop}%
\bibitem [{\citenamefont {Thompson}\ \emph {et~al.}(1992)\citenamefont
  {Thompson}, \citenamefont {Rempe},\ and\ \citenamefont
  {Kimble}}]{Thompson1992}%
  \BibitemOpen
  \bibfield  {author} {\bibinfo {author} {\bibfnamefont {R.~J.}\ \bibnamefont
  {Thompson}}, \bibinfo {author} {\bibfnamefont {G.}~\bibnamefont {Rempe}}, \
  and\ \bibinfo {author} {\bibfnamefont {H.~J.}\ \bibnamefont {Kimble}},\
  }\bibfield  {title} {\enquote {\bibinfo {title} {{Observation of normal-mode
  splitting for an atom in an optical cavity}},}\ }\href {\doibase
  10.1103/PhysRevLett.68.1132} {\bibfield  {journal} {\bibinfo  {journal}
  {Phys. Rev. Lett.}\ }\textbf {\bibinfo {volume} {68}},\ \bibinfo {pages}
  {1132} (\bibinfo {year} {1992})}\BibitemShut {NoStop}%
\bibitem [{\citenamefont {Birnbaum}\ \emph {et~al.}(2005)\citenamefont
  {Birnbaum}, \citenamefont {Boca}, \citenamefont {Miller}, \citenamefont
  {Boozer}, \citenamefont {Northup},\ and\ \citenamefont
  {Kimble}}]{Birnbaum2005}%
  \BibitemOpen
  \bibfield  {author} {\bibinfo {author} {\bibfnamefont {K.~M.}\ \bibnamefont
  {Birnbaum}}, \bibinfo {author} {\bibfnamefont {A.}~\bibnamefont {Boca}},
  \bibinfo {author} {\bibfnamefont {R.}~\bibnamefont {Miller}}, \bibinfo
  {author} {\bibfnamefont {A.~D.}\ \bibnamefont {Boozer}}, \bibinfo {author}
  {\bibfnamefont {T.~E.}\ \bibnamefont {Northup}}, \ and\ \bibinfo {author}
  {\bibfnamefont {H.~J.}\ \bibnamefont {Kimble}},\ }\bibfield  {title}
  {\enquote {\bibinfo {title} {{Photon blockade in an optical cavity with one
  trapped atom.}}}\ }\href {\doibase 10.1038/nature03804} {\bibfield  {journal}
  {\bibinfo  {journal} {Nature}\ }\textbf {\bibinfo {volume} {436}},\ \bibinfo
  {pages} {87} (\bibinfo {year} {2005})}\BibitemShut {NoStop}%
\bibitem [{\citenamefont {Zhang}\ \emph {et~al.}(2011)\citenamefont {Zhang},
  \citenamefont {Guo}, \citenamefont {Li}, \citenamefont {Zhang}, \citenamefont
  {Zhang}, \citenamefont {Du}, \citenamefont {Li}, \citenamefont {Wang},\ and\
  \citenamefont {Zhang}}]{Zhang2011}%
  \BibitemOpen
  \bibfield  {author} {\bibinfo {author} {\bibfnamefont {P.}~\bibnamefont
  {Zhang}}, \bibinfo {author} {\bibfnamefont {Y.}~\bibnamefont {Guo}}, \bibinfo
  {author} {\bibfnamefont {Z.}~\bibnamefont {Li}}, \bibinfo {author}
  {\bibfnamefont {Y.}~\bibnamefont {Zhang}}, \bibinfo {author} {\bibfnamefont
  {Y.}~\bibnamefont {Zhang}}, \bibinfo {author} {\bibfnamefont
  {J.}~\bibnamefont {Du}}, \bibinfo {author} {\bibfnamefont {G.}~\bibnamefont
  {Li}}, \bibinfo {author} {\bibfnamefont {J.}~\bibnamefont {Wang}}, \ and\
  \bibinfo {author} {\bibfnamefont {T.}~\bibnamefont {Zhang}},\ }\bibfield
  {title} {\enquote {\bibinfo {title} {{Elimination of the degenerate
  trajectory of a single atom strongly coupled to a tilted TEM10 cavity
  mode}},}\ }\href {\doibase 10.1103/PhysRevA.83.031804} {\bibfield  {journal}
  {\bibinfo  {journal} {Phys. Rev. A}\ }\textbf {\bibinfo {volume} {83}},\
  \bibinfo {pages} {031804} (\bibinfo {year} {2011})}\BibitemShut {NoStop}%
\bibitem [{\citenamefont {Hacker}\ \emph {et~al.}(2016)\citenamefont {Hacker},
  \citenamefont {Welte}, \citenamefont {Rempe},\ and\ \citenamefont
  {Ritter}}]{Hacker2016}%
  \BibitemOpen
  \bibfield  {author} {\bibinfo {author} {\bibfnamefont {B.}~\bibnamefont
  {Hacker}}, \bibinfo {author} {\bibfnamefont {S.}~\bibnamefont {Welte}},
  \bibinfo {author} {\bibfnamefont {G.}~\bibnamefont {Rempe}}, \ and\ \bibinfo
  {author} {\bibfnamefont {S.}~\bibnamefont {Ritter}},\ }\bibfield  {title}
  {\enquote {\bibinfo {title} {{A photon-photon quantum gate based on a single
  atom in an optical resonator}},}\ }\href {\doibase 10.1038/nature18592}
  {\bibfield  {journal} {\bibinfo  {journal} {Nature}\ }\textbf {\bibinfo
  {volume} {536}},\ \bibinfo {pages} {193} (\bibinfo {year}
  {2016})}\BibitemShut {NoStop}%
\bibitem [{\citenamefont {Raizen}\ \emph {et~al.}(1989)\citenamefont {Raizen},
  \citenamefont {Thompson}, \citenamefont {Brecha}, \citenamefont {Kimble},\
  and\ \citenamefont {Carmichael}}]{Raizen1989}%
  \BibitemOpen
  \bibfield  {author} {\bibinfo {author} {\bibfnamefont {M.~G.}\ \bibnamefont
  {Raizen}}, \bibinfo {author} {\bibfnamefont {R.~J.}\ \bibnamefont
  {Thompson}}, \bibinfo {author} {\bibfnamefont {R.~J.}\ \bibnamefont
  {Brecha}}, \bibinfo {author} {\bibfnamefont {H.~J.}\ \bibnamefont {Kimble}},
  \ and\ \bibinfo {author} {\bibfnamefont {H.~J.}\ \bibnamefont {Carmichael}},\
  }\bibfield  {title} {\enquote {\bibinfo {title} {{Normal-mode splitting and
  linewidth averaging for two-state atoms in an optical cavity}},}\ }\href
  {\doibase 10.1103/PhysRevLett.63.240} {\bibfield  {journal} {\bibinfo
  {journal} {Phys. Rev. Lett.}\ }\textbf {\bibinfo {volume} {63}},\ \bibinfo
  {pages} {240} (\bibinfo {year} {1989})}\BibitemShut {NoStop}%
\bibitem [{\citenamefont {Zhu}\ \emph {et~al.}(1990)\citenamefont {Zhu},
  \citenamefont {Gauthier}, \citenamefont {Morin}, \citenamefont {Wu},
  \citenamefont {Carmichael},\ and\ \citenamefont {Mossberg}}]{Zhu1990}%
  \BibitemOpen
  \bibfield  {author} {\bibinfo {author} {\bibfnamefont {Y.}~\bibnamefont
  {Zhu}}, \bibinfo {author} {\bibfnamefont {D.~J.}\ \bibnamefont {Gauthier}},
  \bibinfo {author} {\bibfnamefont {S.~E.}\ \bibnamefont {Morin}}, \bibinfo
  {author} {\bibfnamefont {Q.}~\bibnamefont {Wu}}, \bibinfo {author}
  {\bibfnamefont {H.~J.}\ \bibnamefont {Carmichael}}, \ and\ \bibinfo {author}
  {\bibfnamefont {T.~W.}\ \bibnamefont {Mossberg}},\ }\bibfield  {title}
  {\enquote {\bibinfo {title} {{Vacuum Rabi splitting as a feature of
  linear-dispersion theory: Analysis and experimental observations}},}\ }\href
  {\doibase 10.1103/PhysRevLett.64.2499} {\bibfield  {journal} {\bibinfo
  {journal} {Phys. Rev. Lett.}\ }\textbf {\bibinfo {volume} {64}},\ \bibinfo
  {pages} {2499} (\bibinfo {year} {1990})}\BibitemShut {NoStop}%
\bibitem [{\citenamefont {{Min Xiao}}(2003)}]{MinXiao2003}%
  \BibitemOpen
  \bibfield  {author} {\bibinfo {author} {\bibnamefont {{Min Xiao}}},\
  }\bibfield  {title} {\enquote {\bibinfo {title} {{Novel linear and nonlinear
  optical properties of electromagnetically induced transparency systems}},}\
  }\href {\doibase 10.1109/JSTQE.2002.807973} {\bibfield  {journal} {\bibinfo
  {journal} {IEEE J. Sel. Top. Quantum Electron.}\ }\textbf {\bibinfo {volume}
  {9}},\ \bibinfo {pages} {86} (\bibinfo {year} {2003})}\BibitemShut {NoStop}%
\bibitem [{\citenamefont {Tanji-Suzuki}\ \emph {et~al.}(2011)\citenamefont
  {Tanji-Suzuki}, \citenamefont {Leroux}, \citenamefont {Schleier-Smith},
  \citenamefont {Cetina}, \citenamefont {Grier}, \citenamefont {Simon},\ and\
  \citenamefont {Vuleti{\'{c}}}}]{Tanji-Suzuki2011}%
  \BibitemOpen
  \bibfield  {author} {\bibinfo {author} {\bibfnamefont {H.}~\bibnamefont
  {Tanji-Suzuki}}, \bibinfo {author} {\bibfnamefont {I.~D.}\ \bibnamefont
  {Leroux}}, \bibinfo {author} {\bibfnamefont {M.~H.}\ \bibnamefont
  {Schleier-Smith}}, \bibinfo {author} {\bibfnamefont {M.}~\bibnamefont
  {Cetina}}, \bibinfo {author} {\bibfnamefont {A.~T.}\ \bibnamefont {Grier}},
  \bibinfo {author} {\bibfnamefont {J.}~\bibnamefont {Simon}}, \ and\ \bibinfo
  {author} {\bibfnamefont {V.}~\bibnamefont {Vuleti{\'{c}}}},\ }\href {\doibase
  10.1016/B978-0-12-385508-4.00004-8} {\emph {\bibinfo {title} {Adv. At. Mol.
  Opt. Phys.}}},\ Vol.~\bibinfo {volume} {60}\ (\bibinfo {year} {2011})\ pp.\
  \bibinfo {pages} {201--237}\BibitemShut {NoStop}%
\bibitem [{\citenamefont {Jing}\ \emph {et~al.}(2014)\citenamefont {Jing},
  \citenamefont {Zhou}, \citenamefont {Liu}, \citenamefont {Qin}, \citenamefont
  {Fang}, \citenamefont {Zhou},\ and\ \citenamefont {Zhang}}]{Jing2014}%
  \BibitemOpen
  \bibfield  {author} {\bibinfo {author} {\bibfnamefont {J.}~\bibnamefont
  {Jing}}, \bibinfo {author} {\bibfnamefont {Z.}~\bibnamefont {Zhou}}, \bibinfo
  {author} {\bibfnamefont {C.}~\bibnamefont {Liu}}, \bibinfo {author}
  {\bibfnamefont {Z.}~\bibnamefont {Qin}}, \bibinfo {author} {\bibfnamefont
  {Y.}~\bibnamefont {Fang}}, \bibinfo {author} {\bibfnamefont {J.}~\bibnamefont
  {Zhou}}, \ and\ \bibinfo {author} {\bibfnamefont {W.}~\bibnamefont {Zhang}},\
  }\bibfield  {title} {\enquote {\bibinfo {title} {{Ultralow-light-level
  all-optical transistor in rubidium vapor}},}\ }\href {\doibase
  10.1063/1.4871384} {\bibfield  {journal} {\bibinfo  {journal} {Appl. Phys.
  Lett.}\ }\textbf {\bibinfo {volume} {104}},\ \bibinfo {pages} {151103}
  (\bibinfo {year} {2014})}\BibitemShut {NoStop}%
\bibitem [{\citenamefont {Simon}\ \emph {et~al.}(2007)\citenamefont {Simon},
  \citenamefont {Tanji}, \citenamefont {Thompson},\ and\ \citenamefont
  {Vuleti{\'{c}}}}]{Simon2007}%
  \BibitemOpen
  \bibfield  {author} {\bibinfo {author} {\bibfnamefont {J.}~\bibnamefont
  {Simon}}, \bibinfo {author} {\bibfnamefont {H.}~\bibnamefont {Tanji}},
  \bibinfo {author} {\bibfnamefont {J.~K.}\ \bibnamefont {Thompson}}, \ and\
  \bibinfo {author} {\bibfnamefont {V.}~\bibnamefont {Vuleti{\'{c}}}},\
  }\bibfield  {title} {\enquote {\bibinfo {title} {{Interfacing Collective
  Atomic Excitations and Single Photons}},}\ }\href {\doibase
  10.1103/PhysRevLett.98.183601} {\bibfield  {journal} {\bibinfo  {journal}
  {Phys. Rev. Lett.}\ }\textbf {\bibinfo {volume} {98}},\ \bibinfo {pages}
  {183601} (\bibinfo {year} {2007})}\BibitemShut {NoStop}%
\bibitem [{\citenamefont {Wu}\ \emph {et~al.}(2008)\citenamefont {Wu},
  \citenamefont {Gea-Banacloche},\ and\ \citenamefont {Xiao}}]{Wu2008}%
  \BibitemOpen
  \bibfield  {author} {\bibinfo {author} {\bibfnamefont {H.}~\bibnamefont
  {Wu}}, \bibinfo {author} {\bibfnamefont {J.}~\bibnamefont {Gea-Banacloche}},
  \ and\ \bibinfo {author} {\bibfnamefont {M.}~\bibnamefont {Xiao}},\
  }\bibfield  {title} {\enquote {\bibinfo {title} {{Observation of Intracavity
  Electromagnetically Induced Transparency and Polariton Resonances in a
  Doppler-Broadened Medium}},}\ }\href {\doibase
  10.1103/PhysRevLett.100.173602} {\bibfield  {journal} {\bibinfo  {journal}
  {Phys. Rev. Lett.}\ }\textbf {\bibinfo {volume} {100}},\ \bibinfo {pages}
  {173602} (\bibinfo {year} {2008})}\BibitemShut {NoStop}%
\bibitem [{\citenamefont {Hernandez}\ \emph {et~al.}(2009)\citenamefont
  {Hernandez}, \citenamefont {Zhang},\ and\ \citenamefont
  {Zhu}}]{Hernandez2009}%
  \BibitemOpen
  \bibfield  {author} {\bibinfo {author} {\bibfnamefont {G.}~\bibnamefont
  {Hernandez}}, \bibinfo {author} {\bibfnamefont {J.}~\bibnamefont {Zhang}}, \
  and\ \bibinfo {author} {\bibfnamefont {Y.}~\bibnamefont {Zhu}},\ }\bibfield
  {title} {\enquote {\bibinfo {title} {{Collective coupling of atoms with
  cavity mode and free-space field}},}\ }\href {\doibase 10.1364/OE.17.004798}
  {\bibfield  {journal} {\bibinfo  {journal} {Opt. Express}\ }\textbf {\bibinfo
  {volume} {17}},\ \bibinfo {pages} {4798} (\bibinfo {year}
  {2009})}\BibitemShut {NoStop}%
\bibitem [{\citenamefont {Fang-Yen}\ \emph {et~al.}(2006)\citenamefont
  {Fang-Yen}, \citenamefont {Yu}, \citenamefont {Ha}, \citenamefont {Choi},
  \citenamefont {An}, \citenamefont {Dasari},\ and\ \citenamefont
  {Feld}}]{Fang-Yen2006}%
  \BibitemOpen
  \bibfield  {author} {\bibinfo {author} {\bibfnamefont {C.}~\bibnamefont
  {Fang-Yen}}, \bibinfo {author} {\bibfnamefont {C.~C.}\ \bibnamefont {Yu}},
  \bibinfo {author} {\bibfnamefont {S.}~\bibnamefont {Ha}}, \bibinfo {author}
  {\bibfnamefont {W.}~\bibnamefont {Choi}}, \bibinfo {author} {\bibfnamefont
  {K.}~\bibnamefont {An}}, \bibinfo {author} {\bibfnamefont {R.~R.}\
  \bibnamefont {Dasari}}, \ and\ \bibinfo {author} {\bibfnamefont {M.~S.}\
  \bibnamefont {Feld}},\ }\bibfield  {title} {\enquote {\bibinfo {title}
  {{Observation of multiple thresholds in the many-atom cavity QED
  microlaser}},}\ }\href {\doibase 10.1103/PhysRevA.73.041802} {\bibfield
  {journal} {\bibinfo  {journal} {Phys. Rev. A}\ }\textbf {\bibinfo {volume}
  {73}},\ \bibinfo {pages} {041802} (\bibinfo {year} {2006})}\BibitemShut
  {NoStop}%
\bibitem [{\citenamefont {Guerlin}\ \emph {et~al.}(2010)\citenamefont
  {Guerlin}, \citenamefont {Brion}, \citenamefont {Esslinger},\ and\
  \citenamefont {M{\o}lmer}}]{Guerlin2010}%
  \BibitemOpen
  \bibfield  {author} {\bibinfo {author} {\bibfnamefont {C.}~\bibnamefont
  {Guerlin}}, \bibinfo {author} {\bibfnamefont {E.}~\bibnamefont {Brion}},
  \bibinfo {author} {\bibfnamefont {T.}~\bibnamefont {Esslinger}}, \ and\
  \bibinfo {author} {\bibfnamefont {K.}~\bibnamefont {M{\o}lmer}},\ }\bibfield
  {title} {\enquote {\bibinfo {title} {{Cavity quantum electrodynamics with a
  Rydberg-blocked atomic ensemble}},}\ }\href {\doibase
  10.1103/PhysRevA.82.053832} {\bibfield  {journal} {\bibinfo  {journal} {Phys.
  Rev. A}\ }\textbf {\bibinfo {volume} {82}},\ \bibinfo {pages} {053832}
  (\bibinfo {year} {2010})}\BibitemShut {NoStop}%
\bibitem [{\citenamefont {Yu}\ \emph {et~al.}(2009)\citenamefont {Yu},
  \citenamefont {Xiong}, \citenamefont {Chen}, \citenamefont {Wang},
  \citenamefont {Xiao},\ and\ \citenamefont {Zhang}}]{Yu2009}%
  \BibitemOpen
  \bibfield  {author} {\bibinfo {author} {\bibfnamefont {X.}~\bibnamefont
  {Yu}}, \bibinfo {author} {\bibfnamefont {D.}~\bibnamefont {Xiong}}, \bibinfo
  {author} {\bibfnamefont {H.}~\bibnamefont {Chen}}, \bibinfo {author}
  {\bibfnamefont {P.}~\bibnamefont {Wang}}, \bibinfo {author} {\bibfnamefont
  {M.}~\bibnamefont {Xiao}}, \ and\ \bibinfo {author} {\bibfnamefont
  {J.}~\bibnamefont {Zhang}},\ }\bibfield  {title} {\enquote {\bibinfo {title}
  {{Multi-normal-mode splitting of a cavity in the presence of atoms: A step
  towards the superstrong-coupling regime}},}\ }\href {\doibase
  10.1103/PhysRevA.79.061803} {\bibfield  {journal} {\bibinfo  {journal} {Phys.
  Rev. A}\ }\textbf {\bibinfo {volume} {79}},\ \bibinfo {pages} {061803}
  (\bibinfo {year} {2009})}\BibitemShut {NoStop}%
\bibitem [{\citenamefont {Yu}\ \emph {et~al.}(2010)\citenamefont {Yu},
  \citenamefont {Xiao},\ and\ \citenamefont {Zhang}}]{Yu2010}%
  \BibitemOpen
  \bibfield  {author} {\bibinfo {author} {\bibfnamefont {X.}~\bibnamefont
  {Yu}}, \bibinfo {author} {\bibfnamefont {M.}~\bibnamefont {Xiao}}, \ and\
  \bibinfo {author} {\bibfnamefont {J.}~\bibnamefont {Zhang}},\ }\bibfield
  {title} {\enquote {\bibinfo {title} {{Triply-resonant optical parametric
  oscillator by four-wave mixing with rubidium vapor inside an optical
  cavity}},}\ }\href {\doibase 10.1063/1.3295694} {\bibfield  {journal}
  {\bibinfo  {journal} {Appl. Phys. Lett.}\ }\textbf {\bibinfo {volume} {96}},\
  \bibinfo {pages} {041101} (\bibinfo {year} {2010})}\BibitemShut {NoStop}%
\bibitem [{\citenamefont {Wang}\ \emph {et~al.}(2002)\citenamefont {Wang},
  \citenamefont {Goorskey},\ and\ \citenamefont {Xiao}}]{Wang2002}%
  \BibitemOpen
  \bibfield  {author} {\bibinfo {author} {\bibfnamefont {H.}~\bibnamefont
  {Wang}}, \bibinfo {author} {\bibfnamefont {D.}~\bibnamefont {Goorskey}}, \
  and\ \bibinfo {author} {\bibfnamefont {M.}~\bibnamefont {Xiao}},\ }\bibfield
  {title} {\enquote {\bibinfo {title} {{Controlling light by light with
  three-level atoms inside an optical cavity}},}\ }\href {\doibase
  10.1364/OL.27.001354} {\bibfield  {journal} {\bibinfo  {journal} {Opt.
  Lett.}\ }\textbf {\bibinfo {volume} {27}},\ \bibinfo {pages} {1354} (\bibinfo
  {year} {2002})}\BibitemShut {NoStop}%
\bibitem [{\citenamefont {Sawant}\ and\ \citenamefont
  {Rangwala}(2016)}]{Sawant2016}%
  \BibitemOpen
  \bibfield  {author} {\bibinfo {author} {\bibfnamefont {R.}~\bibnamefont
  {Sawant}}\ and\ \bibinfo {author} {\bibfnamefont {S.~A.}\ \bibnamefont
  {Rangwala}},\ }\bibfield  {title} {\enquote {\bibinfo {title}
  {{Optical-bistability-enabled control of resonant light transmission for an
  atom-cavity system}},}\ }\href {\doibase 10.1103/PhysRevA.93.023806}
  {\bibfield  {journal} {\bibinfo  {journal} {Phys. Rev. A}\ }\textbf {\bibinfo
  {volume} {93}},\ \bibinfo {pages} {023806} (\bibinfo {year}
  {2016})}\BibitemShut {NoStop}%
\bibitem [{\citenamefont {Joshi}\ and\ \citenamefont {Xiao}(2003)}]{Joshi2003}%
  \BibitemOpen
  \bibfield  {author} {\bibinfo {author} {\bibfnamefont {A.}~\bibnamefont
  {Joshi}}\ and\ \bibinfo {author} {\bibfnamefont {M.}~\bibnamefont {Xiao}},\
  }\bibfield  {title} {\enquote {\bibinfo {title} {{Optical Multistability in
  Three-Level Atoms inside an Optical Ring Cavity}},}\ }\href {\doibase
  10.1103/PhysRevLett.91.143904} {\bibfield  {journal} {\bibinfo  {journal}
  {Phys. Rev. Lett.}\ }\textbf {\bibinfo {volume} {91}},\ \bibinfo {pages}
  {143904} (\bibinfo {year} {2003})}\BibitemShut {NoStop}%
\bibitem [{\citenamefont {Dawes}(2005)}]{Dawes2005}%
  \BibitemOpen
  \bibfield  {author} {\bibinfo {author} {\bibfnamefont {A.~M.~C.}\
  \bibnamefont {Dawes}},\ }\bibfield  {title} {\enquote {\bibinfo {title}
  {{All-Optical Switching in Rubidium Vapor}},}\ }\href {\doibase
  10.1126/science.1110151} {\bibfield  {journal} {\bibinfo  {journal}
  {Science}\ }\textbf {\bibinfo {volume} {308}},\ \bibinfo {pages} {672}
  (\bibinfo {year} {2005})}\BibitemShut {NoStop}%
\bibitem [{\citenamefont {Wei}\ \emph {et~al.}(2010)\citenamefont {Wei},
  \citenamefont {Zhang},\ and\ \citenamefont {Zhu}}]{Wei2010}%
  \BibitemOpen
  \bibfield  {author} {\bibinfo {author} {\bibfnamefont {X.}~\bibnamefont
  {Wei}}, \bibinfo {author} {\bibfnamefont {J.}~\bibnamefont {Zhang}}, \ and\
  \bibinfo {author} {\bibfnamefont {Y.}~\bibnamefont {Zhu}},\ }\bibfield
  {title} {\enquote {\bibinfo {title} {{All-optical switching in a coupled
  cavity-atom system}},}\ }\href {\doibase 10.1103/PhysRevA.82.033808}
  {\bibfield  {journal} {\bibinfo  {journal} {Phys. Rev. A}\ }\textbf {\bibinfo
  {volume} {82}},\ \bibinfo {pages} {033808} (\bibinfo {year}
  {2010})}\BibitemShut {NoStop}%
\bibitem [{\citenamefont {Dutta}\ and\ \citenamefont
  {Rangwala}(2017)}]{Dutta2017}%
  \BibitemOpen
  \bibfield  {author} {\bibinfo {author} {\bibfnamefont {S.}~\bibnamefont
  {Dutta}}\ and\ \bibinfo {author} {\bibfnamefont {S.~A.}\ \bibnamefont
  {Rangwala}},\ }\bibfield  {title} {\enquote {\bibinfo {title} {{All-optical
  switching in a continuously operated and strongly coupled atom-cavity
  system}},}\ }\href {\doibase 10.1063/1.4978933} {\bibfield  {journal}
  {\bibinfo  {journal} {Appl. Phys. Lett.}\ }\textbf {\bibinfo {volume}
  {110}},\ \bibinfo {pages} {121107} (\bibinfo {year} {2017})}\BibitemShut
  {NoStop}%
\bibitem [{\citenamefont {Wang}\ \emph {et~al.}(2017)\citenamefont {Wang},
  \citenamefont {Di}, \citenamefont {Zhu},\ and\ \citenamefont
  {Agarwal}}]{Wang2017}%
  \BibitemOpen
  \bibfield  {author} {\bibinfo {author} {\bibfnamefont {L.}~\bibnamefont
  {Wang}}, \bibinfo {author} {\bibfnamefont {K.}~\bibnamefont {Di}}, \bibinfo
  {author} {\bibfnamefont {Y.}~\bibnamefont {Zhu}}, \ and\ \bibinfo {author}
  {\bibfnamefont {G.~S.}\ \bibnamefont {Agarwal}},\ }\bibfield  {title}
  {\enquote {\bibinfo {title} {{Interference control of perfect photon
  absorption in cavity quantum electrodynamics}},}\ }\href {\doibase
  10.1103/PhysRevA.95.013841} {\bibfield  {journal} {\bibinfo  {journal} {Phys.
  Rev. A}\ }\textbf {\bibinfo {volume} {95}},\ \bibinfo {pages} {013841}
  (\bibinfo {year} {2017})}\BibitemShut {NoStop}%
\bibitem [{\citenamefont {Hosseini}\ \emph {et~al.}(2011)\citenamefont
  {Hosseini}, \citenamefont {Sparkes}, \citenamefont {Campbell}, \citenamefont
  {Lam},\ and\ \citenamefont {Buchler}}]{Hosseini2011}%
  \BibitemOpen
  \bibfield  {author} {\bibinfo {author} {\bibfnamefont {M.}~\bibnamefont
  {Hosseini}}, \bibinfo {author} {\bibfnamefont {B.~M.}\ \bibnamefont
  {Sparkes}}, \bibinfo {author} {\bibfnamefont {G.}~\bibnamefont {Campbell}},
  \bibinfo {author} {\bibfnamefont {P.~K.}\ \bibnamefont {Lam}}, \ and\
  \bibinfo {author} {\bibfnamefont {B.~C.}\ \bibnamefont {Buchler}},\
  }\bibfield  {title} {\enquote {\bibinfo {title} {{High efficiency coherent
  optical memory with warm rubidium vapour.}}}\ }\href {\doibase
  10.1038/ncomms1175} {\bibfield  {journal} {\bibinfo  {journal} {Nat.
  Commun.}\ }\textbf {\bibinfo {volume} {2}},\ \bibinfo {pages} {174} (\bibinfo
  {year} {2011})}\BibitemShut {NoStop}%
\bibitem [{\citenamefont {El-Ganainy}\ \emph {et~al.}(2018)\citenamefont
  {El-Ganainy}, \citenamefont {Makris}, \citenamefont {Khajavikhan},
  \citenamefont {Musslimani}, \citenamefont {Rotter},\ and\ \citenamefont
  {Christodoulides}}]{El-Ganainy2018}%
  \BibitemOpen
  \bibfield  {author} {\bibinfo {author} {\bibfnamefont {R.}~\bibnamefont
  {El-Ganainy}}, \bibinfo {author} {\bibfnamefont {K.~G.}\ \bibnamefont
  {Makris}}, \bibinfo {author} {\bibfnamefont {M.}~\bibnamefont {Khajavikhan}},
  \bibinfo {author} {\bibfnamefont {Z.~H.}\ \bibnamefont {Musslimani}},
  \bibinfo {author} {\bibfnamefont {S.}~\bibnamefont {Rotter}}, \ and\ \bibinfo
  {author} {\bibfnamefont {D.~N.}\ \bibnamefont {Christodoulides}},\ }\bibfield
   {title} {\enquote {\bibinfo {title} {{Non-Hermitian physics and PT
  symmetry}},}\ }\href {\doibase 10.1038/nphys4323} {\bibfield  {journal}
  {\bibinfo  {journal} {Nat. Phys.}\ }\textbf {\bibinfo {volume} {14}},\
  \bibinfo {pages} {11} (\bibinfo {year} {2018})}\BibitemShut {NoStop}%
\bibitem [{\citenamefont {Siddons}\ \emph {et~al.}(2008)\citenamefont
  {Siddons}, \citenamefont {Adams}, \citenamefont {Ge},\ and\ \citenamefont
  {Hughes}}]{Siddons2008}%
  \BibitemOpen
  \bibfield  {author} {\bibinfo {author} {\bibfnamefont {P.}~\bibnamefont
  {Siddons}}, \bibinfo {author} {\bibfnamefont {C.~S.}\ \bibnamefont {Adams}},
  \bibinfo {author} {\bibfnamefont {C.}~\bibnamefont {Ge}}, \ and\ \bibinfo
  {author} {\bibfnamefont {I.~G.}\ \bibnamefont {Hughes}},\ }\bibfield  {title}
  {\enquote {\bibinfo {title} {{Absolute absorption on rubidium D lines:
  comparison between theory and experiment}},}\ }\href {\doibase
  10.1088/0953-4075/41/15/155004} {\bibfield  {journal} {\bibinfo  {journal}
  {J. Phys. B At. Mol. Opt. Phys.}\ }\textbf {\bibinfo {volume} {41}},\
  \bibinfo {pages} {155004} (\bibinfo {year} {2008})}\BibitemShut {NoStop}%
\bibitem [{\citenamefont {Weis}\ \emph {et~al.}(2010)\citenamefont {Weis},
  \citenamefont {Riviere}, \citenamefont {Deleglise}, \citenamefont {Gavartin},
  \citenamefont {Arcizet}, \citenamefont {Schliesser},\ and\ \citenamefont
  {Kippenberg}}]{Weis2010}%
  \BibitemOpen
  \bibfield  {author} {\bibinfo {author} {\bibfnamefont {S.}~\bibnamefont
  {Weis}}, \bibinfo {author} {\bibfnamefont {R.}~\bibnamefont {Riviere}},
  \bibinfo {author} {\bibfnamefont {S.}~\bibnamefont {Deleglise}}, \bibinfo
  {author} {\bibfnamefont {E.}~\bibnamefont {Gavartin}}, \bibinfo {author}
  {\bibfnamefont {O.}~\bibnamefont {Arcizet}}, \bibinfo {author} {\bibfnamefont
  {A.}~\bibnamefont {Schliesser}}, \ and\ \bibinfo {author} {\bibfnamefont
  {T.~J.}\ \bibnamefont {Kippenberg}},\ }\bibfield  {title} {\enquote {\bibinfo
  {title} {{Optomechanically Induced Transparency}},}\ }\href {\doibase
  10.1126/science.1195596} {\bibfield  {journal} {\bibinfo  {journal}
  {Science}\ }\textbf {\bibinfo {volume} {330}},\ \bibinfo {pages} {1520}
  (\bibinfo {year} {2010})}\BibitemShut {NoStop}%
\bibitem [{\citenamefont {Singh}\ \emph {et~al.}(2014)\citenamefont {Singh},
  \citenamefont {Bosman}, \citenamefont {Schneider}, \citenamefont {Blanter},
  \citenamefont {Castellanos-Gomez},\ and\ \citenamefont {Steele}}]{Singh2014}%
  \BibitemOpen
  \bibfield  {author} {\bibinfo {author} {\bibfnamefont {V.}~\bibnamefont
  {Singh}}, \bibinfo {author} {\bibfnamefont {S.~J.}\ \bibnamefont {Bosman}},
  \bibinfo {author} {\bibfnamefont {B.~H.}\ \bibnamefont {Schneider}}, \bibinfo
  {author} {\bibfnamefont {Y.~M.}\ \bibnamefont {Blanter}}, \bibinfo {author}
  {\bibfnamefont {A.}~\bibnamefont {Castellanos-Gomez}}, \ and\ \bibinfo
  {author} {\bibfnamefont {G.~a.}\ \bibnamefont {Steele}},\ }\bibfield  {title}
  {\enquote {\bibinfo {title} {{Optomechanical coupling between a multilayer
  graphene mechanical resonator and a superconducting microwave cavity}},}\
  }\href {\doibase 10.1038/nnano.2014.168} {\bibfield  {journal} {\bibinfo
  {journal} {Nat. Nanotechnol.}\ }\textbf {\bibinfo {volume} {9}},\ \bibinfo
  {pages} {820} (\bibinfo {year} {2014})}\BibitemShut {NoStop}%
\bibitem [{\citenamefont {Zhang}\ \emph {et~al.}(2014)\citenamefont {Zhang},
  \citenamefont {Zou}, \citenamefont {Jiang},\ and\ \citenamefont
  {Tang}}]{Zhang2014}%
  \BibitemOpen
  \bibfield  {author} {\bibinfo {author} {\bibfnamefont {X.}~\bibnamefont
  {Zhang}}, \bibinfo {author} {\bibfnamefont {C.-l.}\ \bibnamefont {Zou}},
  \bibinfo {author} {\bibfnamefont {L.}~\bibnamefont {Jiang}}, \ and\ \bibinfo
  {author} {\bibfnamefont {H.~X.}\ \bibnamefont {Tang}},\ }\bibfield  {title}
  {\enquote {\bibinfo {title} {{Strongly coupled magnons and cavity microwave
  photons}},}\ }\href {\doibase 10.1103/PhysRevLett.113.156401} {\bibfield
  {journal} {\bibinfo  {journal} {Phys. Rev. Lett.}\ }\textbf {\bibinfo
  {volume} {113}},\ \bibinfo {pages} {156401} (\bibinfo {year}
  {2014})}\BibitemShut {NoStop}%
\bibitem [{\citenamefont {Fan}\ \emph {et~al.}(2015)\citenamefont {Fan},
  \citenamefont {Fong}, \citenamefont {Poot},\ and\ \citenamefont
  {Tang}}]{Fan2015}%
  \BibitemOpen
  \bibfield  {author} {\bibinfo {author} {\bibfnamefont {L.}~\bibnamefont
  {Fan}}, \bibinfo {author} {\bibfnamefont {K.~Y.}\ \bibnamefont {Fong}},
  \bibinfo {author} {\bibfnamefont {M.}~\bibnamefont {Poot}}, \ and\ \bibinfo
  {author} {\bibfnamefont {H.~X.}\ \bibnamefont {Tang}},\ }\bibfield  {title}
  {\enquote {\bibinfo {title} {{Cascaded optical transparency in
  multimode-cavity optomechanical systems}},}\ }\href {\doibase
  10.1038/ncomms6850} {\bibfield  {journal} {\bibinfo  {journal} {Nat.
  Commun.}\ }\textbf {\bibinfo {volume} {6}},\ \bibinfo {pages} {5850}
  (\bibinfo {year} {2015})}\BibitemShut {NoStop}%
\bibitem [{\citenamefont {Balram}\ \emph {et~al.}(2016)\citenamefont {Balram},
  \citenamefont {Davan{\c{c}}o}, \citenamefont {Song},\ and\ \citenamefont
  {Srinivasan}}]{Balram2016}%
  \BibitemOpen
  \bibfield  {author} {\bibinfo {author} {\bibfnamefont {K.~C.}\ \bibnamefont
  {Balram}}, \bibinfo {author} {\bibfnamefont {M.~I.}\ \bibnamefont
  {Davan{\c{c}}o}}, \bibinfo {author} {\bibfnamefont {J.~D.}\ \bibnamefont
  {Song}}, \ and\ \bibinfo {author} {\bibfnamefont {K.}~\bibnamefont
  {Srinivasan}},\ }\bibfield  {title} {\enquote {\bibinfo {title} {{Coherent
  coupling between radiofrequency, optical and acoustic waves in
  piezo-optomechanical circuits}},}\ }\href {\doibase 10.1038/nphoton.2016.46}
  {\bibfield  {journal} {\bibinfo  {journal} {Nat. Photonics}\ }\textbf
  {\bibinfo {volume} {10}},\ \bibinfo {pages} {346} (\bibinfo {year}
  {2016})}\BibitemShut {NoStop}%
\bibitem [{\citenamefont {Olson}\ and\ \citenamefont
  {Mayer}(2009)}]{Olson2009}%
  \BibitemOpen
  \bibfield  {author} {\bibinfo {author} {\bibfnamefont {A.~J.}\ \bibnamefont
  {Olson}}\ and\ \bibinfo {author} {\bibfnamefont {S.~K.}\ \bibnamefont
  {Mayer}},\ }\bibfield  {title} {\enquote {\bibinfo {title}
  {{Electromagnetically induced transparency in rubidium}},}\ }\href {\doibase
  10.1119/1.3028309} {\bibfield  {journal} {\bibinfo  {journal} {Am. J. Phys.}\
  }\textbf {\bibinfo {volume} {77}},\ \bibinfo {pages} {116} (\bibinfo {year}
  {2009})}\BibitemShut {NoStop}%
\bibitem [{\citenamefont {Ritter}\ \emph {et~al.}(2016)\citenamefont {Ritter},
  \citenamefont {Gruhler}, \citenamefont {Pernice}, \citenamefont
  {K{\"{u}}bler}, \citenamefont {Pfau},\ and\ \citenamefont
  {L{\"{o}}w}}]{Ritter2016}%
  \BibitemOpen
  \bibfield  {author} {\bibinfo {author} {\bibfnamefont {R.}~\bibnamefont
  {Ritter}}, \bibinfo {author} {\bibfnamefont {N.}~\bibnamefont {Gruhler}},
  \bibinfo {author} {\bibfnamefont {W.~H.~P.}\ \bibnamefont {Pernice}},
  \bibinfo {author} {\bibfnamefont {H.}~\bibnamefont {K{\"{u}}bler}}, \bibinfo
  {author} {\bibfnamefont {T.}~\bibnamefont {Pfau}}, \ and\ \bibinfo {author}
  {\bibfnamefont {R.}~\bibnamefont {L{\"{o}}w}},\ }\bibfield  {title} {\enquote
  {\bibinfo {title} {{Coupling thermal atomic vapor to an integrated ring
  resonator}},}\ }\href {\doibase 10.1088/1367-2630/18/10/103031} {\bibfield
  {journal} {\bibinfo  {journal} {New J. Phys.}\ }\textbf {\bibinfo {volume}
  {18}},\ \bibinfo {pages} {103031} (\bibinfo {year} {2016})}\BibitemShut
  {NoStop}%
\bibitem [{\citenamefont {Stern}\ \emph {et~al.}(2016)\citenamefont {Stern},
  \citenamefont {Zektzer}, \citenamefont {Mazurski},\ and\ \citenamefont
  {Levy}}]{Stern2016}%
  \BibitemOpen
  \bibfield  {author} {\bibinfo {author} {\bibfnamefont {L.}~\bibnamefont
  {Stern}}, \bibinfo {author} {\bibfnamefont {R.}~\bibnamefont {Zektzer}},
  \bibinfo {author} {\bibfnamefont {N.}~\bibnamefont {Mazurski}}, \ and\
  \bibinfo {author} {\bibfnamefont {U.}~\bibnamefont {Levy}},\ }\bibfield
  {title} {\enquote {\bibinfo {title} {{Enhanced light-vapor interactions and
  all optical switching in a chip scale micro-ring resonator coupled with
  atomic vapor}},}\ }\href {\doibase 10.1002/lpor.201600176} {\bibfield
  {journal} {\bibinfo  {journal} {Laser Photon. Rev.}\ }\textbf {\bibinfo
  {volume} {10}},\ \bibinfo {pages} {1016} (\bibinfo {year}
  {2016})}\BibitemShut {NoStop}%
\bibitem [{\citenamefont {Stern}\ \emph {et~al.}(2017)\citenamefont {Stern},
  \citenamefont {Desiatov}, \citenamefont {Mazurski},\ and\ \citenamefont
  {Levy}}]{Stern2017}%
  \BibitemOpen
  \bibfield  {author} {\bibinfo {author} {\bibfnamefont {L.}~\bibnamefont
  {Stern}}, \bibinfo {author} {\bibfnamefont {B.}~\bibnamefont {Desiatov}},
  \bibinfo {author} {\bibfnamefont {N.}~\bibnamefont {Mazurski}}, \ and\
  \bibinfo {author} {\bibfnamefont {U.}~\bibnamefont {Levy}},\ }\bibfield
  {title} {\enquote {\bibinfo {title} {{Strong coupling and high-contrast
  all-optical modulation in atomic cladding waveguides}},}\ }\href {\doibase
  10.1038/ncomms14461} {\bibfield  {journal} {\bibinfo  {journal} {Nat.
  Commun.}\ }\textbf {\bibinfo {volume} {8}},\ \bibinfo {pages} {14461}
  (\bibinfo {year} {2017})}\BibitemShut {NoStop}%
\end{thebibliography}

\end{document}